\documentclass{aastex62}

\renewcommand{\thefootnote}{}

\usepackage[T1]{fontenc}
\usepackage{ae,aecompl}
\usepackage{natbib} 

\submitjournal{ApJ}

\begin{document}

\title{Discovery of an accretion-rate independent absolute RMS amplitude of millihertz quasi-periodic oscillations in 4U 1636$-$53}

\author{Ming Lyu}
\affiliation{Department of Physics, Xiangtan University, Xiangtan, Hunan 411105, China}

\author{Mariano M\'endez}
\affiliation{Kapteyn Astronomical Institute, University of Groningen, PO BOX 800, NL-9700 AV Groningen, the Netherlands}

\author{D. Altamirano}
\affiliation{Physics \& Astronomy, University of Southampton, Southampton, Hampshire SO17 1BJ, UK}

\author{Guobao Zhang}
\affiliation{Yunnan Observatories, Chinese Academy of Sciences (CAS), Kunming 650216, P.R. China}
\affiliation{Key Laboratory for the Structure and Evolution of Celestial Objects, CAS, Kunming 650216, P.R. China}

\author{G. C. Mancuso}
\affiliation{Instituto Argentino de Radioastronom\'{\i}a (CCT-La Plata, CONICET; CICPBA), C.C. No. 5, 1894 Villa Elisa, Argentina}
\affiliation{Facultad de Ciencias Astron\'omicas y Geof\'{\i}sicas, Universidad Nacional de La Plata, Paseo del Bosque s/n, 1900 La Plata, Argentina}
\affiliation{Physics \& Astronomy, University of Southampton, Southampton, Hampshire SO17 1BJ, UK}


\email{lvming@xtu.edu.cn}



\begin{abstract}
We investigate the frequency and amplitude of the millihertz quasi-periodic oscillations (mHz QPOs) in the neutron-star low-mass X-ray binary (NS LMXB) 4U 1636--53 using Rossi X-ray Timing Explorer observations. We find that no mHz QPOs appear when the source is in the hard spectral state. We also find that there is no significant correlation between the frequency and the fractional rms amplitude of the mHz QPOs. Notwithstanding, for the first time, we find that the absolute RMS amplitude of the mHz QPOs is insensitive to the parameter S$_{a}$, which measures the position of the source in the colour-colour diagram and is usually assumed to be an increasing function of mass accretion rate. This finding indicates that the transition from marginally stable burning to stable burning or unstable burning could happen very rapidly since, before the transition, the mHz QPOs do not gradually decay as the rate further changes.

\end{abstract}



\section{Introduction}

A unique class of quasi-periodic oscillations (QPOs) was firstly discovered by \citet{revni01} in three neutron-star low-mass X-ray binaries (LMXBs), 4U 1608$-$52, 4U 1636$-$53, and Aql X$-$1. The unique properties of these QPOs are \citep{revni01,diego08,lyu15}: (1) Their typical frequency range is very low, about 6-14 mHz; (2) these QPOs are stronger at low photon energies ($<$ 5 keV); (3) they appear only when the source is within a narrow range of X-ray luminosities, $L_{\rm 2-20\:keV} \simeq (5-11) \times 10^{36}$ ergs s$^{-1}$; (4) they disappear or become undetectable when there is a type I X-ray burst.

\citet{revni01} proposed that the mHz QPOs originate from a special mode of nuclear burning on the neutron star surface, when the mass accretion rate is within a certain range. \citet{heger07} proposed that the mHz QPOs could be a consequence of marginally stable nuclear burning of Helium on the neutron-star surface. In the model of \citet{heger07}, the characteristic time scale of the oscillations is $\sim$100 seconds, remarkably consistent with the $\sim$2-minute period of the mHz QPOs. Besides, the model predicts that the oscillations should occur only in a very narrow range of X-ray luminosity. However, the marginally stable nuclear burning in the model occurs only when the accretion rate is close to the Eddington rate, one order of magnitude higher than the averaged global rate over the entire neutron star surface, calculated from the X-ray luminosity at which mHz QPOs are observed. \citet{heger07} proposed that the local accretion rate in the burning layer where the QPOs happen can be higher than the global accretion rate.

An anti-correlation between the frequency of the kilohertz (kHz) QPO and the 2--5 keV X-ray count rate associated with a 7.5 mHz QPO has been reported in 4U 1608--52 \citep{yu02}. This result supported the nuclear burning interpretation of the mHz QPOs: as the luminosity increases, the stresses of the radiation from the neutron-star surface push outwards the inner disc in each mHz QPO cycle, thus leading to the change of kHz QPO frequency.

\citet{diego08} found that, in the transitional state between the soft and hard spectral state, the frequency of the mHz QPO in 4U 1636$-$53 decreased systematically with time, until the oscillations disappear and a type I X-ray burst occurs. Very recently, \citet{Mancuso19} found a similar behavior in the LMXB EXO 0748--676. This frequency drift behaviour further supported the idea that mHz QPOs are closely connected to nuclear burning on the neutron star surface. The discovery of `high-luminosity' mHz QPOs in the neutron star transient source IGR J17480$-$2446 \citep{linaries10,chakraborty11} indicated that some mHz QPOs appear with different characteristics. The `high-luminosity' QPOs have a frequency of about 4.5 mHz, and the persistent luminosity of this source when the mHz QPOs were observed was $L_{2-50{\rm keV}}$ $\sim$ 10$^{38}$ erg s$^{-1}$. \citet{linaries12} showed that there is a smooth transition between mHz QPO and thermonuclear bursts in IGR J17480$-$2446: as the accretion rate increased, bursts gradually evolved into a mHz QPO, and vice versa. This evolution is consistent with the prediction from the marginally stable burning model of \citet{heger07}, further supporting the idea that mHz QPOs are due to marginally stable burning on the neutron star surface.

\citet{stiele16} studied phase-resolved energy spectra of the mHz QPOs in 4U 1636$-$53 and found that the oscillations were not caused by variations in the blackbody temperature of the neutron star surface. Conversely, \citet{strohmayer18} recently found that the mHz oscillations in the `clocked burster' GS 1826$-$238 were consistent with being produced by modulation of the temperature component of the blackbody emission from the neutron star surface, assuming a constant blackbody normalization throughout the oscillation cycle. This finding favours the model by \citet{heger07}, however, as emphasized by \citet{strohmayer18}, given the current data, that is not the only possible interpretation.
Therefore, the connection between the mHz oscillations and the variation of the thermal radiation from the neutron star surface remains an open question.

A scenario considering the turbulent chemical mixing of the fuel, together with a higher heat flux from the crust, is able to explain the observed accretion rate at which mHz QPOs are seen \citep{keek09}. Furthermore, \citet{keek09} found that the frequency drift of the QPOs before X-ray bursts may be due to the cooling process of the layer where the mHz QPOs occur. \citet{keek14} investigated the influence of the fuel composition and nuclear reaction rates on the mHz QPOs, and concluded that no allowed variation in the composition and the reaction rate is able to trigger the mHz QPOs at the observed accretion rates.

\citet{lyu15} found that there was no significant correlation between the frequency of the mHz QPOs and the temperature of the neutron-star surface in 4U 1636$-$53, which is different from theoretical predictions. Furthermore, \citet{lyu15} found that seven X-ray bursts associated with mHz QPOs in this source were bright, energetic and short, indicating a potential connection between the mHz QPOs and He-rich X-ray bursts. \citet{lyu16} investigated the convexity of 39 type I X-ray bursts associated with mHz QPOs in 4U 1636$-$53 and found that all the bursts show positive convexity. This finding suggests that these mHz QPOs and the associated bursts may originate at the equator of the neutron star surface.

In this work we investigate the properties of mHz QPOs detected with the Rossi X-ray Timing Explorer (RXTE) in the LMXB 4U 1636$-$53 and, for the first time, we carry out a systematic study of the frequency and the amplitude of the mHz QPO. Furthermore, we explore the connection between the frequency, the amplitude, the count rate and the position of the source in the colour-colour diagram, which provides useful information to study the origin of the mHz QPOs. In Section 2 we describe the observations and data analysis used in this work, and in Section 3 we present our main results. Finally, we discuss the results in the framework of the marginally stable nuclear burning model in Section 4.

\section{Observations and data reduction}
We analysed the observations with mHz QPOs reported in \citet{lyu16} in 4U 1636$-$53 using the Proportional Counter Array \citep[PCA;][]{jahoda06} on board of RXTE. An RXTE observation typically covers 1 to 5 consecutive 90-minute satellite orbits, which usually contains between 1 and 5 ks of useful data separated by 1-4 ks data gaps. We extracted two event mode PCA light curves of 1-s resolution in the $\sim 2-4.5$ keV range \citep[where the mHz QPOs are the strongest, see][]{diego08}, with one light curve from all available Proportional Counter Units (PCUs) and the other one from PCU2 only. Bursts and instrumental dropouts were removed.

Since mHz QPOs are not always present in the whole observation, we further divided each light curve extracted from all available PCUs into a series of independent 700 seconds intervals, and used Lomb-Scargle periodograms \citep{lomb76,scargle82} to check whether there is a significant QPO in each interval. We found that, in some cases, although there is a more than $3\sigma$ significant QPO in an observation, the QPO is less than $3\sigma$ significant in some of the 700-s intervals in that observation. This can either happen because the QPO is indeed not present during those intervals, or because the amplitude of the QPO or the source count rate decreased slightly in that interval and the QPO dropped below the $3\sigma$ detection level. Besides, in some cases, random fluctuations may marginally reach the $3\sigma$ significance level, likely where the count rate increases. Therefore, we consider only those intervals where either: (i) There are at least 2 consecutive 700-s intervals in which the QPO is more than 3$\sigma$ significant; or (ii) if there is only one 700-s interval and the QPO there is more than $3\sigma$ significant, then there must be a harmonic at twice the QPO frequency.

\begin{figure}
\center
\includegraphics[height=0.4\textwidth]{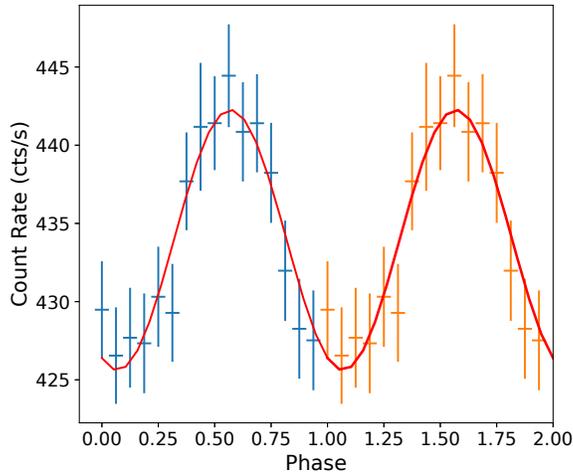}
\caption{A fitting example of the folded light curve extracted from an RXTE/PCA data interval (14000s-14700s in dc7fff0-dc867e5 in 50030-02-10-00) of 4U 1636--53. The red line represents the best fitting result with the model consisting of a constant plus a sine function. Here we show the data and the model curve for two periods for clarity purpose.}
\label{fits}
\end{figure}

For each 700s interval with mHz QPO, we then folded the corresponding 700-s light curve from all available PCUs at the period derived from the frequency measured in the Lomb-Scargle periodogram. We took the frequency at which the power is the strongest in the Lomb-Scargle periodogram to be the frequency of the mHz QPO in each interval. We used the half width at half maximum as an indication of the error of the frequency of the QPOs. We used the ftool $\bold{efold}$ to fold the light curve and normalized the folded light curve to counts/s, with the error bars in the folded light curve evaluated by error propagation. We then fitted each folded light curve with a constant plus a sine function to measure the fractional rms amplitude of the mHz QPO (see Figure \ref{fits}, for example). To estimate the contribution of background to the total count rate, we used the tool $\bold{pcabackest}$ to create an estimated PCA  background light curve in the $\sim 2-4.5$ keV and calculated the average count rate. We then calculated the fractional rms amplitude of the mHz QPO using the formula, $rms=A /[\sqrt{2}*(C-B)]$, where A is the amplitude of the sine function, C is the value of the constant component and B is the count rate of the estimated background. Furthermore, we calculated the absolute RMS amplitude of the mHz QPOs detected by PCU2 detector, $RMS=rms \times R$, where R is the count rate from the PCU2 detector.

To explore the possible link between the properties of the mHz QPOs and the different accretion states of the source, we made the colour-colour diagram and further traced the evolution of the frequency and the amplitude of the mHz QPO as function of the quantity S$_{a}$ (see below). We used the 16-s time-resolution Standard-2 data available from the RXTE/PCA to calculate two X-ray colours. We defined the soft colour as the count rate in the 3.5-6.0 keV band divided by the count rate in the 2.0-3.5 keV band, and the hard colour as the count rate in the 9.7-16.0 keV band divided by the count rate in the 6.0-9.7 keV band. The colours are normalised to those of the Crab Nebula in observations taken close in time to the ones of 4U 1636--53 \citep[see][for more details]{diego08b,zhang09}. We defined the S$_{a}$ quantity on the basis of the colour-colour diagram of the source as in \citet{zhang11}, with the S$_{a}$ length being normalised to the distance between S$_{a}$ = 1 at the top right-hand vertex and S$_{a}$ = 2 at the bottom left-hand vertex of the colour-colour diagram. The S$_{a}$ value of each particular observation is then assigned to be the same as the value of the point which is closest to that observation along the S$_{a}$ curve. The quantity S$_{a}$ is usually assumed to be an increasing function of mass accretion rate \citep{hasinger89,zhang11}.

\section{Results}

\begin{figure}
\center
\includegraphics[height=0.4\textwidth]{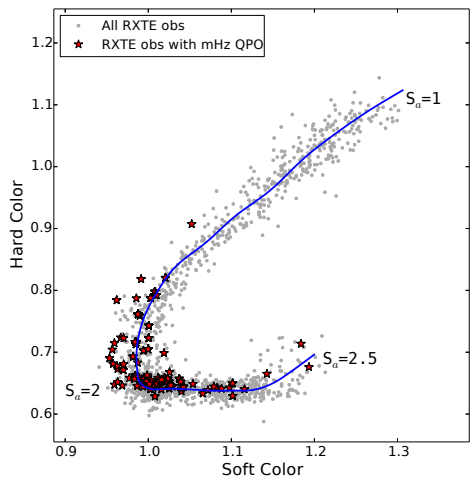}
\caption{Colour-colour diagram of 4U 1636--53 using all RXTE observations. Each gray point represents the averaged Crab-normalised colours \citep[see][for details]{zhang11} of a single RXTE observation. The red stars mark the position of the observations where mHz QPOs are detected. The position of the source in the diagram is parameterised by the length of the blue solid curve S$_{a}$ \citep[see also][]{zhang11}.}
\label{ccd}
\end{figure}

\begin{figure}
\center
\includegraphics[height=0.4\textwidth]{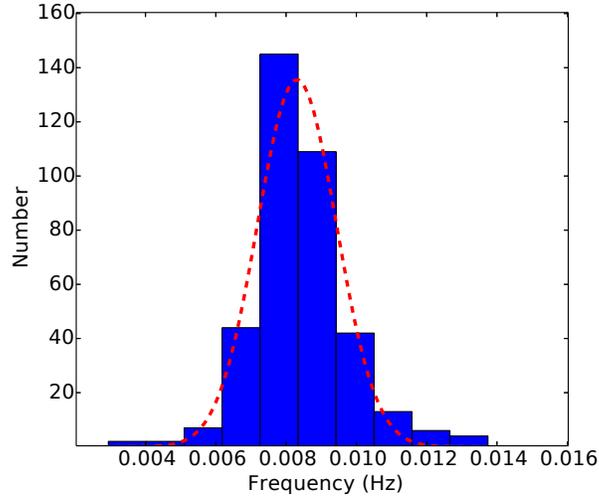}
\caption{Distribution of the frequency of the mHz QPOs in each 700s interval in 4U 1636--53. The red dashed line in the plot corresponds to the best-fitting Gaussian curve to the histogram.}
\label{hist_freq}
\end{figure}

\begin{figure}
\center
\includegraphics[height=0.4\textwidth]{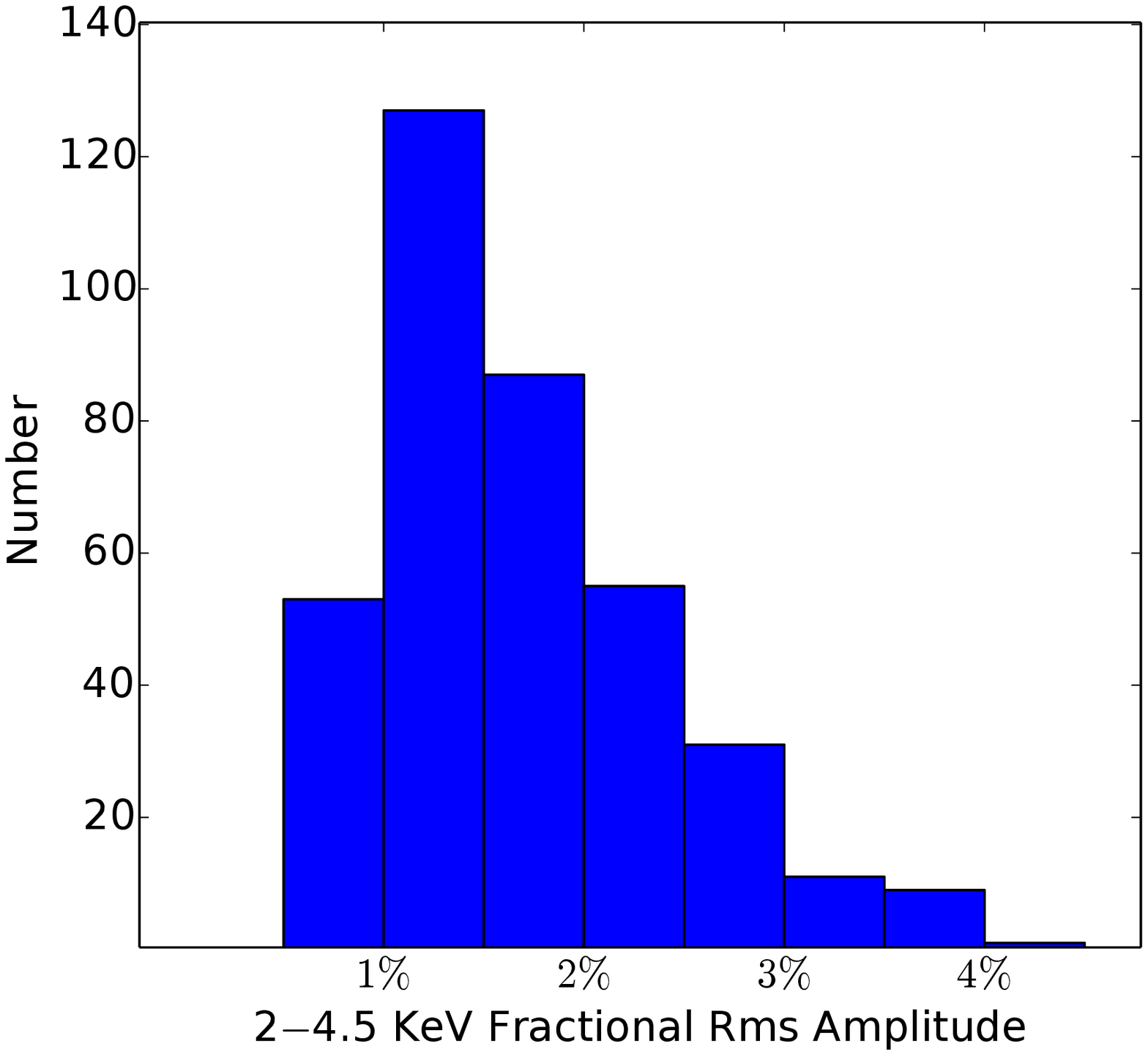}
\includegraphics[height=0.4\textwidth]{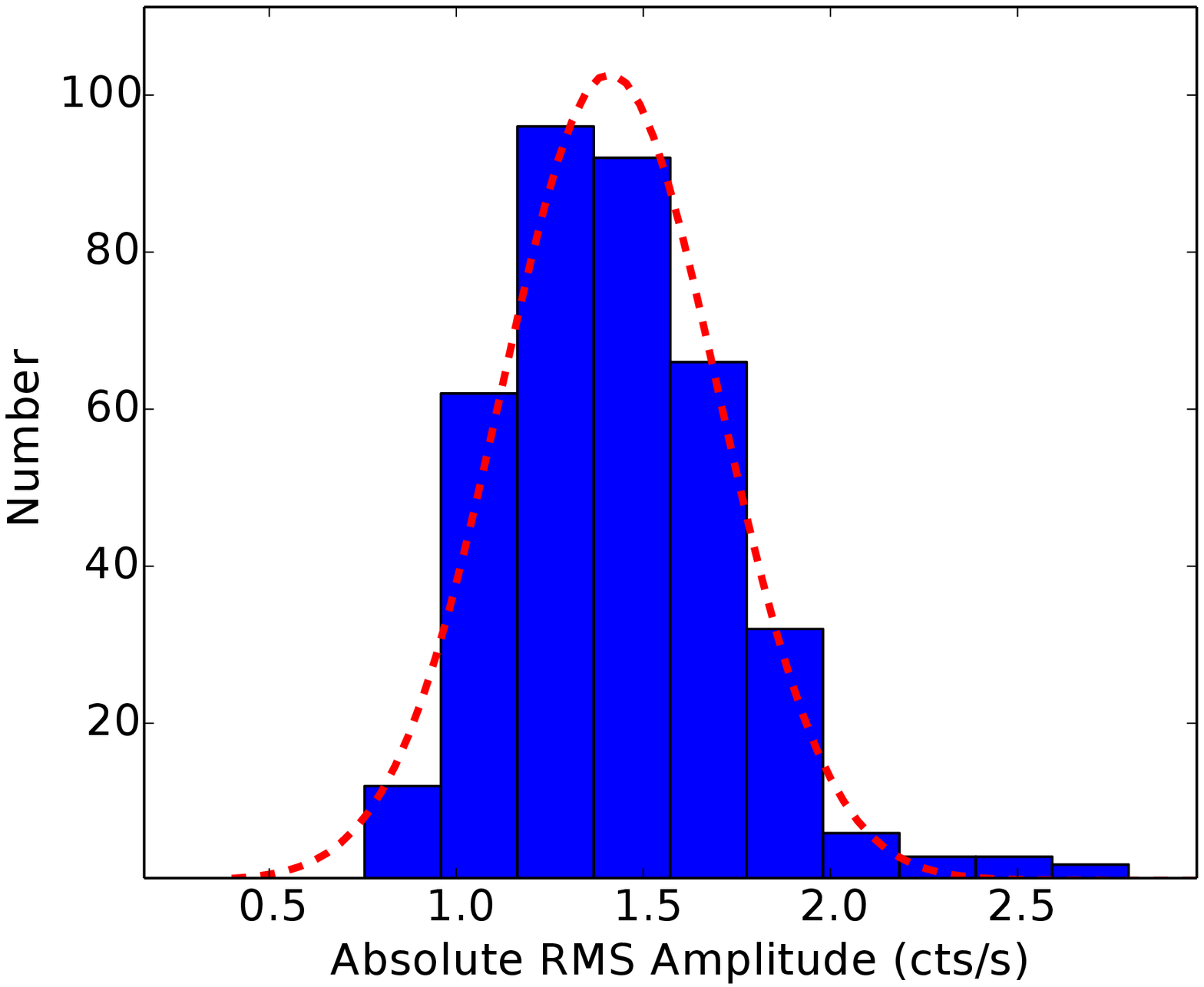}
\caption{Distribution of the fractional rms amplitude and absolute RMS amplitude of the mHz QPOs in each 700s interval in 4U 1636--53. The red dashed line in the right plot corresponds to the best-fitting Gaussian curve to the histogram. The absolute RMS amplitude is measured with PCU2 detector in 2-4.5 keV range.}
\label{hist_rms}
\end{figure}

\begin{figure}
\center
\includegraphics[height=0.4\textwidth]{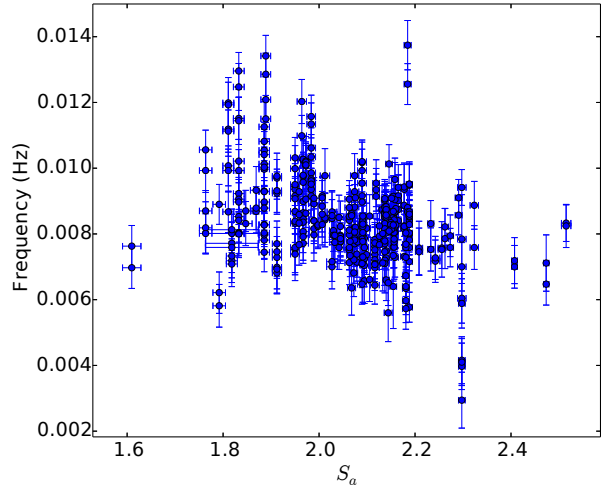}
\caption{Frequency of the mHz QPOs vs. S$_{a}$ in 4U 1636--53. Each data point in the plot corresponds to a 700 seconds interval.}
\label{fre_sa}
\end{figure}

\begin{figure}
\center
\includegraphics[height=0.4\textwidth]{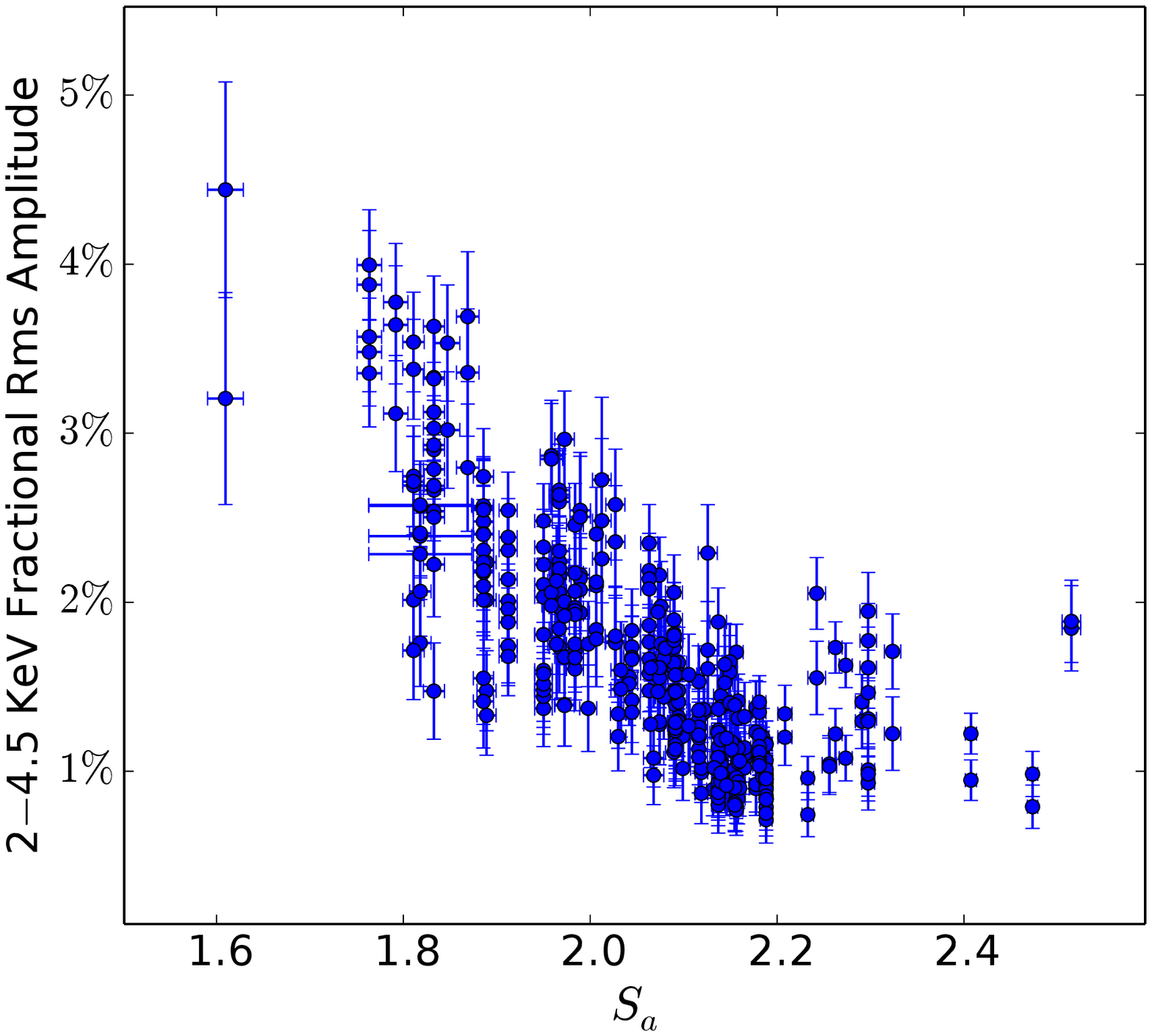}
\includegraphics[height=0.4\textwidth]{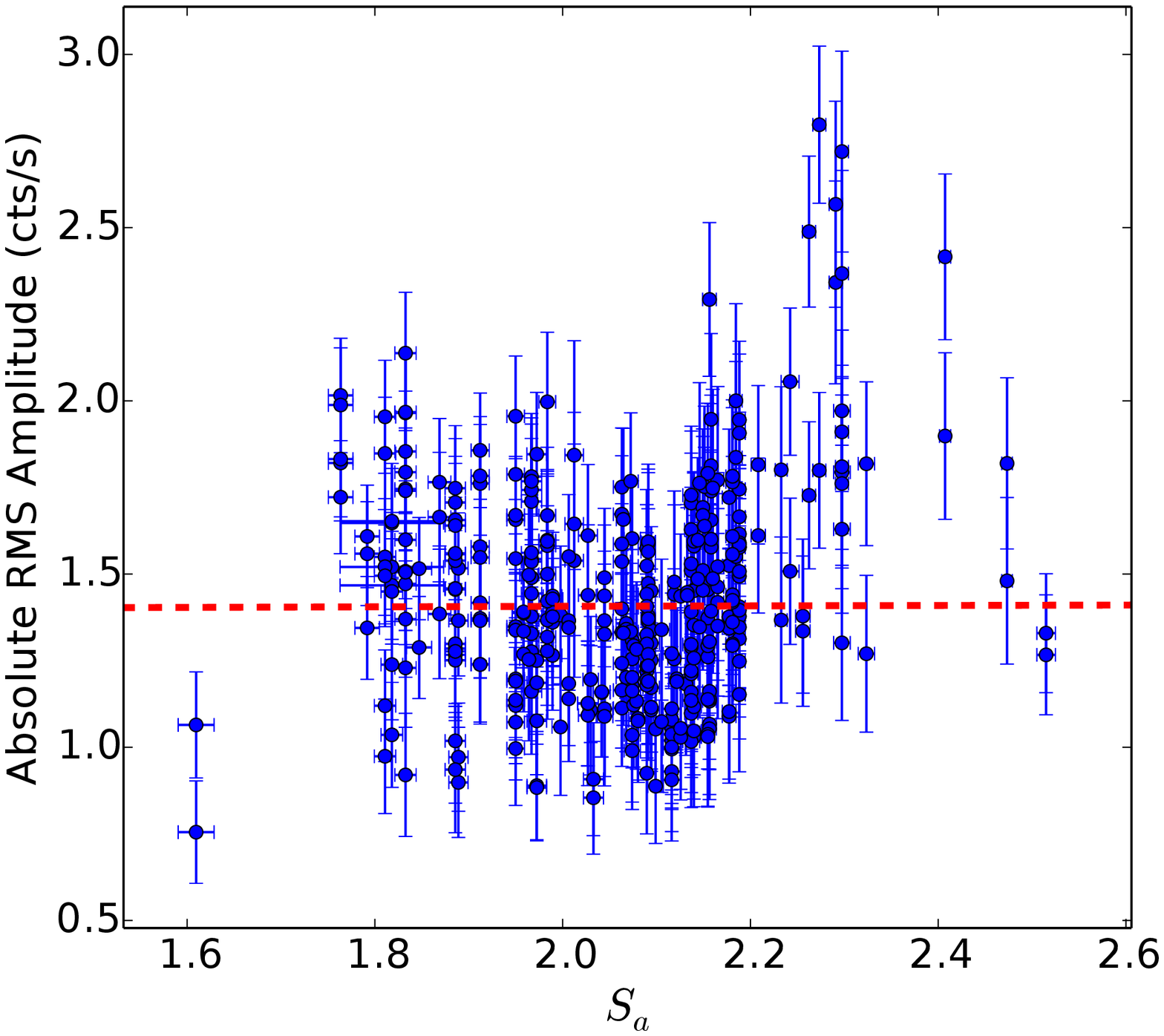}
\caption{Fractional rms and absolute RMS amplitude of mHz QPOs vs. S$_{a}$ in 4U 1636--53. Each data point in the plot corresponds to a 700 seconds interval. The absolute RMS amplitude is measured with PCU2 detector in 2-4.5 keV range.}
\label{rms_sa}
\end{figure}

\begin{figure}
\center
\includegraphics[height=0.4\textwidth]{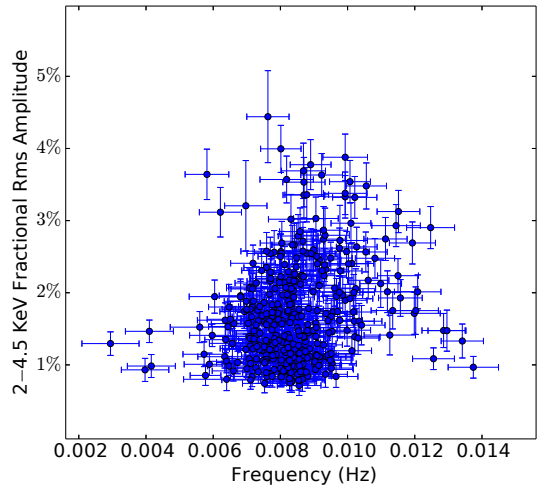}
\caption{Fractional rms amplitude of the mHz QPOs vs. the frequency of the mHz QPOs in 4U 1636--53. Each data point in the plot corresponds to a 700 seconds interval.}
\label{fre_rms}
\end{figure}

\begin{figure}
\center
\includegraphics[height=0.4\textwidth]{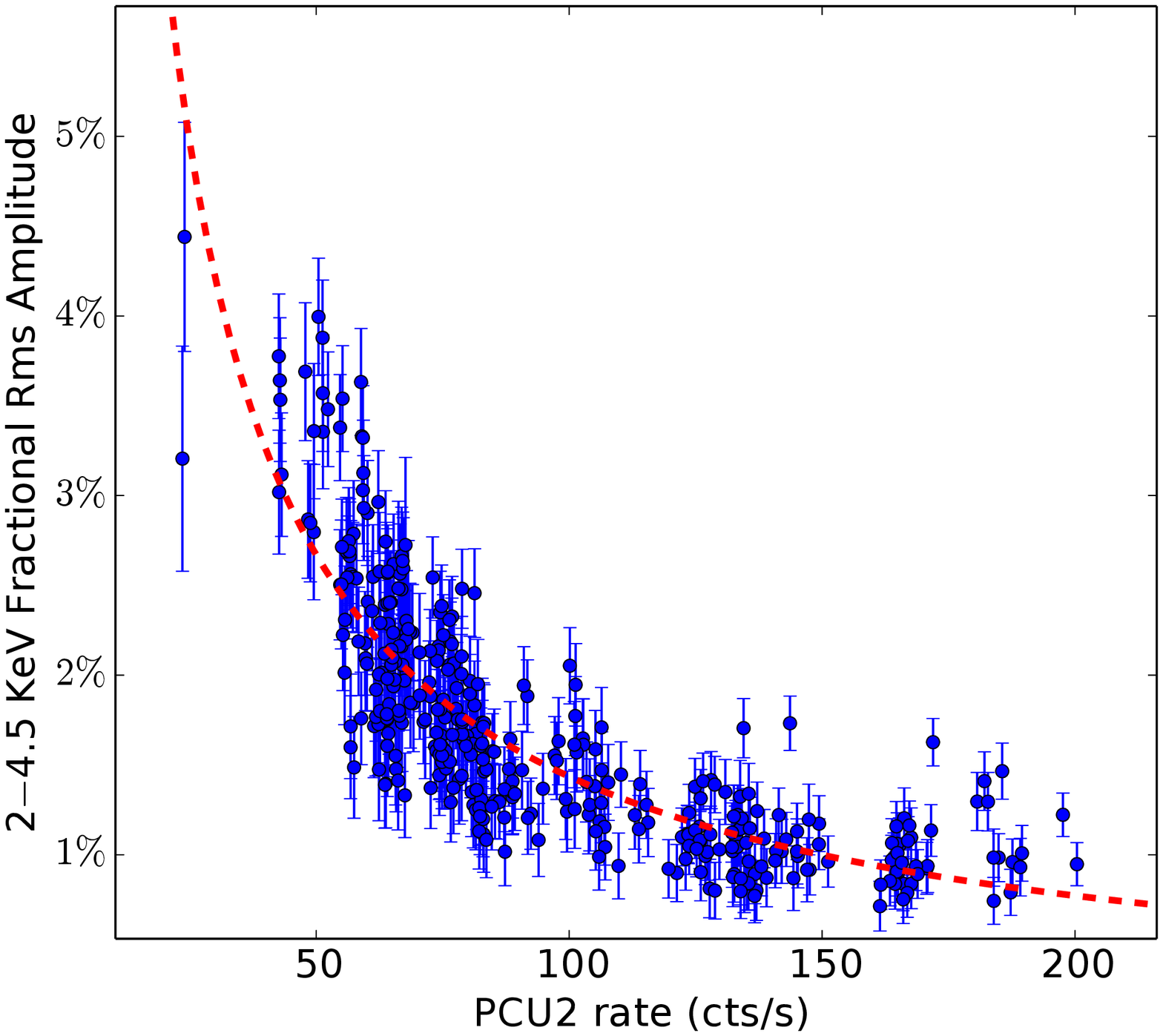}
\caption{Fractional rms amplitude of the mHz QPOs as a function of the count rate of the PCU2 light curve ($<$$\sim$5 keV) in 4U 1636--53. The red dashed lines in the plots correspond to the best-fitting power-law model to the data. Each data point in the plot corresponds to a 700 seconds interval.}
\label{rate_rms}
\end{figure}

In total, we detected mHz QPOs in 374 individual 700-second intervals. In Figure \ref{ccd} we show the colour-colour diagram for all RXTE observations of 4U 1636--53. We find that the observations with mHz QPOs are either in the transitional spectral state or in the soft spectral state, with most of them clustering around the vertex between the two branches, at S$_{a}$$\sim$2. The hard colour of the observations with mHz QPOs is less than $\sim$0.9, suggesting that the mHz QPOs do not appear in the hard spectral state.

In Figure \ref{hist_freq} we show the distribution of the frequency of the mHz QPOs in 4U 1636$-$53. The distribution of the frequency is symmetric, ranging from $\sim$3 mHz to $\sim$14 mHz. As shown in the plot, the distribution can be well described by a Gaussian function with a mean frequency of 8.31$\pm$0.06 mHz (all errors are at 68\% confidence level) and a standard deviation of 1.12$\pm$0.05 mHz. We show the distribution of the fractional rms amplitude and absolute RMS amplitude of the mHz QPOs in Figure \ref{hist_rms}. The fractional rms amplitude ranges from $\sim$0.5\% to $\sim$4.5\%, with most of the values clustering between 0.5\% and 2.5\%. The distribution of the absolute RMS amplitude of the mHz QPOs is symmetric, if we fit a Gauss function to these data the best-fitting average value and the standard deviation are 1.41$\pm$0.02 cts/s and 0.29$\pm$0.01 cts/s, respectively.

In Figure \ref{fre_sa} and Figure \ref{rms_sa} we plot, respectively, the frequency, the fractional and absolute RMS amplitudes of the mHz QPOs in 4U 1636$-$53 as a function of the parameter S$_{a}$. There is no clear correlation between the QPO frequency and S$_{a}$ (correlation coefficient$=-0.4$), with most of the data in the figure clustering in the S$_{a}$ range $2.0-2.2$. The absolute RMS amplitude remains more or less constant as S$_{a}$ increases; if we fit a power-law function to these data the best-fitting powerlaw index is $\sim0.01\pm0.10$. On the contrary, there is a significant anti-correlation between the fractional rms amplitude and S$_{a}$ (correlation coefficient$=-0.8$). The fractional rms amplitude decreases as S$_{a}$ increases up to S$_{a} \sim$ 2.2, and then it remains more or less constant when S$_{a}$ increases above $\sim$2.2.

In Figure \ref{fre_rms} we show the relation between the frequency and the fractional rms amplitude of the mHz QPOs in 4U 1636$-$53. We find no clear correlation between the frequency and the rms amplitude. The correlation coefficient for the data in this plot is $\sim0.2$, indicating that these two quantities are not significantly correlated.

In Figure \ref{rate_rms} we show the fractional rms amplitude of the mHz QPOs in 4U 1636--53 vs. the count rate of the PCU2 detector below $\sim$ 5 keV. We find that there is a significant anti-correlation between the rate and the fractional rms amplitude. As the count rate increases, the fractional rms amplitude drops fast when the count rate is below $\sim$100 cts/s, and then it decreases much more slowly when the count rate is larger than 100 cts/s. The anti-correlation between the rate and the fractional rms amplitude can be well fitted by a power-law relation with a power-law index$^{*}$ $-0.89\pm0.02$.
\footnote{* The index may be slightly overestimated due to the fact that the fractional rms amplitude of the QPOs decreases rapidly at high count rate and, since the significance of a QPO scales with the square of the fractional rms amplitude \citep{vanderklis98}, we tend to miss mHz QPOs that are marginally significant at these count rates in Figure \ref{rate_rms}.}

\section{Discussion}
We systematically investigated the properties of the mHz QPOs in 4U 1636--53 using RXTE observations. We found that the mHz QPOs are present only in the soft spectral state or the transitional state between the soft and the hard state. We found no clear correlation between the frequency and the fractional rms amplitude of the mHz QPOs. More importantly, the absolute RMS amplitude of the mHz QPOs in 4U 1636--53 remains constant as the S$_{a}$ increases, as the source gradually moves to the soft state. Furthermore, we found an anti-correlation between the fractional rms amplitude and the source count rate below $\sim$ 5 keV.

The finding that mHz QPOs are not present in the hard spectral state suggests that the mHz QPOs can be triggered only when the mass accretion rate exceeds a certain threshold value. This is consistent with the model of \citet{heger07}: The mHz QPOs appear only when the mass accretion rate increases above certain value. The critical accretion rate at which mHz QPOs are triggered in \citet{heger07} is very close to the Eddington accretion rate, significantly higher than the averaged global accretion rate deduced from X-ray luminosity. Moreover, the model of \citet{heger07} predicts a very narrow local accretion rate range where mHz QPOs happen, whereas we find the mHz QPOs over a relatively wide S$_{a}$ range, from $\sim$1.6 to 2.5 of the full range of S$_{a}$ values (see Figure \ref{fre_sa} and Figure \ref{rms_sa}). If S$_{a}$ is indeed correlated to $\dot M$ \citep{hasinger89,zhang11}, this result implies that in 4U 1636$-$53 either the mHz QPOs exist over a wide range of $\dot M$, in contradiction to the model, or in the region of the colour-colour diagram in which the mHz QPOs are detected the $\dot M$ is not strongly correlated to S$_{a}$. On the other hand, when the mHz QPOs are present the soft X-ray intensity changes by a factor of $\sim$ 4 (see Figure \ref{rate_rms}), which favours the idea that in 4U 1636$-$53, when the mHz QPOs are present, $\dot M$ changes by a large factor. A scenario which can well bridge the accretion rate difference between models and observations is that the mHz QPOs are triggered by the local accretion rate around equator of the neutron star surface instead of the global accretion rate of the whole neutron-star surface \citep{heger07,diego08,lyu16}.

\citet{revni01} investigated the properties of the mHz QPOs and found that the centroid frequency of the mHz QPOs in the $\sim$ 2-5 keV energy range is $\sim$ $7-9$ mHz. \citet{diego08} found that the frequency of the mHz QPOs ($\sim$ 2-5 keV) in 4U 1636--53 ranges between $\sim$ 7 to 14 mHz. The frequency range of the mHz QPOs derived in this work is $\sim$ 3 to 14 mHz. We measured the average frequency of the mHz QPOs at a much smaller time interval (700 seconds) compared with previous authors, and this may account for the relatively wider frequency range measured in this work. For the same reason, the range of the fractional rms amplitude of the mHz QPOs that we found here is also a bit larger than the fractional rms amplitude range ($\sim$ 0.6\%-2\%) reported in \citet{revni01}. The fractional rms amplitude of the mHz QPOs in this work can reach up to $\sim$ 4\% in 4U 1636--53.  

A linear relationship between the absolute RMS amplitude of short-term variability and flux variations on longer timescales was firstly discovered in X-ray binary systems as well as in active galactic nuclei \citep{uttley01}. Further analysis showed that this linear RMS-flux relation occurs on all measured timescales \citep{uttley05}. This RMS-flux relation in accreting systems is likely linked to the accretion flow itself, with variations arising from fluctuations in the accretion rate at different radii, which propagate through the flow so that variability is coupled together over a broad range of time-scales \citep{lyubar97,king04,arevalo06,ingram13,scaringi14,cow14,hogg16}. In this work we found an anti-correlation between the fractional rms amplitude of the mHz QPOs and the count rate below 5 keV in 4U 1636--53. The derived power-law index ($\sim$-0.9) indicates that there is no significant correlation between the absolute RMS amplitude of the mHz QPOs and the soft count rate in 4U 1636--53, different from the linear RMS-flux relationship described above. This difference indicates that the relation in this work is likely to have a different physical origin compared with the one of the linear relationship described above, further supporting the idea that the mHz QPOs likely originate from the neutron-star surface. However, The physical mechanism behind the relation in this work is still unknown. \citet{keek09} simulated the nuclear burning process on the neutron-star surface and found that oscillations in their simulations generally exhibit larger amplitudes as the heat flux from the neutron star crust decreases. If the heat flux is proportional to the temperature of the neutron-star surface, and hence correlated with the number of soft photons from the neutron-star, then there should be an anti-correlation between the amplitude of the mHz QPOs and the count rate in the soft band.

The independence of the absolute RMS amplitude with the parameter S$_{a}$ suggests that the disappearance of the mHz QPOs should be a very quick process. Simulations in \citet{heger07} indicate that the marginally stable nuclear burning switches to the stable burning/unstable burning as the accretion rate further increases/decreases, leading to the disappearance of the mHz QPOs. If the absolute RMS amplitude remains constant when S$_{a}$ changes, then this switch should be suddenly triggered and then rapidly finished, consistent with the observational feature that the mHz QPOs suddenly disappear when there is a type I burst.

\section{Acknowledgements}
This research has made use of data obtained from the High Energy Astrophysics Science Archive Research Center (HEASARC), provided by NASA's Goddard Space Flight Center. This research made use of NASA's Astrophysics Data System. Lyu is supported by National Natural Science Foundation of China (grant No.11803025); and the Hunan Provincial Natural Science Foundation (grant No. 2018JJ3483) and Hunan Education Department Foundation (grant No. 17C1520). DA acknowledges support from the Royal Society. G.B. acknowledges funding support from the National Natural Science Foundation of China (NSFC) under grant numbers U1838116 and  the CAS Pioneer Hundred Talent Program Y7CZ181002. GCM thanks the Royal Society International Exchanges program for their support. GCM was partially supported by PIP 0102 (CONICET) and received financial support from PICT-2017-2865 (ANPCyT).

\bibliography{paper}



\end{document}